\documentstyle [12pt] {article}

\catcode`@=12
\begin{document}
\thispagestyle{empty}
\begin{flushright} UCRHEP-T160\\August 1996\
\end{flushright}
\vspace{0.5in}
\begin{center}
{\Large	\bf Baryon-Number Nonconservation and\\}
\vspace{0.1in}
{\Large \bf the Stability of Strange Matter\\}
\vspace{1.5in}
{\bf E. Keith and Ernest Ma\\}
\vspace{0.3in}
{\sl Department of Physics\\}
{\sl University of California\\}
{\sl Riverside, California 92521\\}
\vspace{1.5in}
\end{center}

\begin{abstract}\
If baryon-number nonconservation exists in the form of an effective 
six-quark operator which also changes strangeness by four units, it 
could have a tremendous impact on the absolute stability of strange quark 
matter.  We show that such an operator is negligible in the supersymmetric 
standard model with $\lambda_{ijk} u_i^c d_j^c d_k^c$ terms in the 
superpotential, but may be of importance in models with exotic particle 
content.  From the experimental lower limit of $10^{25}$ years on the 
stability of nuclei, we find a model-independent lower limit of the order 
$10^5$ years on the stability of strange matter against such decays.
\end{abstract}

\newpage
\baselineskip 24pt

\section{Introduction}

Matter containing a large number of strange quarks may have a lower energy 
per baryon than ordinary nuclei and be absolutely stable\cite{1,2}. 
This intriguing possibility has generated a great deal of interest across 
many subfields of physics.  A crucial implicit assumption for the stability 
of this new kind of matter, called strange matter, is that the baryon number 
$B$ is exactly conserved.  Of course, we know nuclei are stable against 
$\Delta B \neq 0$ decays.  The best mode-independent experimental lower 
limit to date is $1.6 \times 10^{25}$ years\cite{3}.  However, there may be 
effective $\Delta B \neq 0$ interactions which are highly suppressed for 
nuclei but not for strange matter.  We examine such a hypothesis here and 
show that whereas such exotic decays are possible, the relevant lifetime 
should be longer than $10^5$ years.

In Sec.~2 we review briefly the present experimental constraints on 
$\Delta B \neq 0$ interactions.  These come mainly from the nonobservation 
of proton decay and of neutron-antineutron oscillation.  We then point out 
the possible consequences of an effective $\Delta B = 2$, $\Delta N_s = 4$ 
operator (where $N_s$ denotes the number of $s$ quarks) on the stability of 
strange matter.  We proceed to discuss some possible theoretical origins of 
such an operator in Sec.~3 and Sec.~4.  We deal first with the supersymmetric 
standard model with $R$-parity nonconserving terms of the form 
$\lambda_{ijk} u_i^c d_j^c d_k^c$ and show that their effects are 
negligible.  We then discuss two possible extended models where the 
effects may be large.  In Sec.~5 we consider this 
operator in conjunction with the usual weak interaction and show that 
the suppression coming from the stability of ordinary nuclei translates 
to a phenomenological lower limit of $10^5$ years on the lifetime of 
strange matter.  Finally in Sec.~6, there are some concluding remarks.

\section{Baryon-Number Nonconserving Interactions}

Given the standard $SU(3) \times SU(2) \times U(1)$ gauge symmetry and the 
usual quarks and leptons, it is well-known that 
the resulting renormalizable Lagrangian conserves both the 
baryon number $B$ and the lepton number $L$ automatically. 
However, new physics at higher energy scales may induce effective 
interactions (of dimension 5 or above) which do not respect these 
conservation laws.  The foremost example is the possibility of proton 
decay.  For this to happen, there has to be at least one fermion with a 
mass below that of the proton.  Since only leptons are known to have this 
property, the selection rule $\Delta B = 1$, $\Delta L = 1$ is applicable. 
The most studied decay mode experimentally is $p \rightarrow \pi^0 e^+$, 
which requires the effective interaction
\begin{equation}
{\cal H}_{int} \sim {1 \over M^2} (uude).
\end{equation}
It is now known\cite{4} that $\tau_{exp} (p \rightarrow \pi^0 e^+) > 9 \times 
10^{32}$ years, hence $M > 10^{16}$ GeV.  [Here $M$ should be considered 
an effective mass, because in some scenarios the denominator of the 
right-hand side of Eq.~(1) is really the product of two different masses.] 
This means that proton decay probes physics at the grand-unification 
energy scale.

The next simplest class of effective interactions has the selection rule 
$\Delta B = 2$, $\Delta L = 0$.  The most well-known example is of course
\begin{equation}
{\cal H}_{int} \sim {1 \over M^5} (udd)^2,
\end{equation}
which induces neutron-antineutron oscillation and allows a nucleus to decay 
by the annihilation of two of its nucleons.  Note that since 6 fermions are 
now involved, the effective operator is of dimension 9, hence $M$ appears to 
the power $-5$.  This means that for the same level of nuclear stability, 
the lower bound on $M$ will be only of order $10^5$ GeV.  The present best 
experimental lower limit on the $n - \bar n$ oscillation lifetime is\cite{5} 
$8.6 \times 10^7$s.  Direct search for $N N \rightarrow \pi \pi$ decay in 
iron yields\cite{6} a limit of $6.8 \times 10^{30}$y.  Although the above 
two numbers differ by 30 orders of magnitude, it is well established by 
general arguments as well as detailed nuclear model calculations that\cite{7}
\begin{equation}
\tau (n \bar n) = 8.6 \times 10^7{\rm s} ~ \Rightarrow ~ \tau (N N \rightarrow 
\pi \pi) \sim 2 \times 10^{31}{\rm y}.
\end{equation}
Hence the two limits are comparable.  To estimate the magnitude of $M$, we 
use
\begin{equation}
\tau(n \bar n) = M^5 |\psi(0)|^{-4},
\end{equation}
where the effective wavefunction at the origin is roughly given by\cite{8}
\begin{equation}
|\psi(0)|^{-2} \sim \pi R^3, ~~~ R \sim 1 ~ {\rm fm}.
\end{equation}
Hence we obtain $M > 2.4 \times 10^5$ GeV.  Bearing in mind that 
${\cal H}_{int}$ is likely to be suppressed also by products of couplings 
less than unity, this means that the stability of nuclei is sensitive to 
new physics at an energy scale not too far above the electroweak scale of 
$10^2$ GeV.  It may thus have the hope of future direct experimental 
exploration.

Consider next the effective interaction
\begin{equation}
{\cal H}_{int} \sim {1 \over M^5} (uds)^2.
\end{equation}
This has the selection rule $\Delta B = 2$, $\Delta N_s = 2 ~(\Delta S = -2)$. 
In this case, a nucleus may decay by the process $N N \rightarrow 
K K$\cite{9,10}.  Although such decay modes have never been observed, the 
mode-independent stability lifetime of $1.6 \times 10^{25}$y mentioned 
already is enough to guarantee that it is very small.  However, consider now
\begin{equation}
{\cal H}_{int} \sim {1 \over M^5} (uss)^2,
\end{equation}
which has the selection rule
\begin{equation}
\Delta B = 2, ~~~ \Delta N_s = 4.
\end{equation}
To get rid of four units of strangeness, the nucleus must now convert two 
nucleons into four kaons, but that is kinematically impossible.  Hence the 
severe constraint from the stability of nuclei does not seem to apply here, 
and the following interesting possibility may occur.

Strangelets ({\it i.e.} stable and metastable configurations of quarks 
with large strangeness content) with atomic number $A$ (which is of course 
the same as baryon number $B$) and number of strange quarks $N_s$ may now 
decay into other strangelets with two less units of $A$ and one to three less 
$s$ quarks.  For example,
\begin{eqnarray}
(A, N_s) &\rightarrow& (A - 2, N_s - 2) + K K, \\ 
(A, N_s) &\rightarrow& (A - 2, N_s - 1) + K K K, ~etc.
\end{eqnarray}
For the states of lowest energy, model calculations show\cite{11} that $N_s 
\sim 0.8 A$, hence the above decay modes are efficient ways of reducing all 
would-be stable strangelets to those of the smallest $A$.

Unlike nuclei which are most stable for $A$ near that of iron, the energy 
per baryon number of strangelets decreases with increasing $A$.  This has 
led to the intriguing speculation that there are stable macroscopic lumps 
of strange matter in the Universe.  On the other hand, it is not clear how 
such matter would form, because there are no stable building blocks such 
as hydrogen and helium which are essential for the formation of heavy nuclei. 
In any case, the above exotic interaction would allow strange matter to 
dissipate into smaller and smaller units, until $A$ becomes too small for 
the strangelet itself to be stable.  A sample calculation by Madsen\cite{11} 
shows that
\begin{equation}
m_0 (A) - m_0 (A - 2) \simeq (1704 + 111 A^{-1/3} + 161 A^{-2/3})~{\rm MeV},
\end{equation}
assuming $m_s = 100$ MeV, and the bag factor $B^{1/4} = 145$ MeV.  Hence the 
$KK$ and $KKK$ decays would continue until the ground-state mass $m_0$ 
exceeds the condition that the energy per baryon number is less than 930 
MeV, which happens at around $A = 13$.

\section{Supersymmetric Standard Model}

If the standard model of quarks and leptons is extended to include 
supersymmetry, the $\Delta B \neq 0$ terms $\lambda_{ijk} u_i^c d_j^c d_k^c$ 
are allowed in the superpotential.  In the above notation, all chiral 
superfields are assumed to be left-handed, hence $q^c$ denotes the left-handed 
charge-conjugated quark, or equivalently the right-handed quark, and the 
subscripts refer to families, {\it i.e.} $u_i$ for $(u, c, t)$ and $d_j$ for 
$(d, s, b)$.  Since all quarks are color triplets and the interaction must 
be a singlet which is antisymmetric in color, the two $d$ quark superfields 
must belong to different families.  In the minimal supersymmetric standard 
model, these terms are forbidden by the imposition of $R$-parity.  However, 
this assumption is not mandatory and there is a vast body of recent 
literature exploring the consequences of $R$-parity nonconservation\cite{12}.

Starting with Yukawa terms of the form $\lambda_{ijk} u_i^c d_j^c d_k^c$, 
where two of the fields are quarks and the third is a scalar quark, we can 
obtain the effective interaction of Eq.~(7) in several ways\cite{13}.  In 
Ref.~[10] the $d d \rightarrow \tilde b \tilde b$ box diagram (where 
$\tilde q$ denotes the supersymmetric scalar partner of $q$) is considered to 
obtain the effective interaction of Eq.~(2) for $n - \bar n$ oscillation 
using $\lambda_{udb}$.  Here we take
\begin{equation}
s s \rightarrow \tilde b \tilde b
\end{equation}
and use $\lambda_{usb}$ instead.  The box diagram involves the exchange of 
the $(u,c,t)$ quarks, the $W$ boson, and their supersymmetric partners. 
Since only gauge couplings appear, the Glashow-Iliopoulos-Maiani (GIM) 
mechanism\cite{14} is operative and this diagram vanishes if the 
$(\tilde u,\tilde c,\tilde t)$ scalar quarks have the same mass and that 
the $(u,c,t)$ quark masses can be neglected.  Thus in Fig.~3 of Ref.~[10], 
there is a peak at $M_{\tilde t} = 200$ GeV.  However, there are actually 
two scalar quarks $(\tilde q_L,\tilde q_R)$ for each quark $q$.  Consider 
just $t$ and $(\tilde t_L,\tilde t_R)$.  The $\tilde t$ mass matrix is 
given by
\begin{equation}
{\cal M}^2_{\tilde t} = \left[ \begin{array} {c@{\quad}c} \tilde m_L^2 + 
m_t^2 & A m_t \\ A m_t & \tilde m_R^2 + m_t^2 \end{array} \right].
\end{equation}
Since only $\tilde t_L$ couples to $b_L$ through the gauge fermion $\tilde w$ 
and $\tilde t_L$ is not a mass eigenstate, its approximate effective 
contribution is given by
\begin{equation}
M_{\tilde t_L}^{-2} = {{\tilde m_R^2 + m_t^2} \over {(\tilde m_L^2 + m_t^2) 
(\tilde m_R^2 + m_t^2) - A^2 m_t^2}}.
\end{equation}
Unless this somehow cancels the $\tilde u$ and $\tilde c$ contributions 
accidentally, the box diagram will not vanish.  Furthermore, it is often 
assumed that the soft supersymmetry-breaking terms $\tilde m_L^2$ and 
$\tilde m_R^2$ are universal, in which case this amplitude would be zero 
if the $(u,c,t)$ quark masses were neglected.  Rewriting Eq.~(7) more 
specifically as
\begin{equation}
{\cal H}_{int} = T_1 \epsilon_{\alpha \beta \gamma} \epsilon_{\alpha' 
\beta' \gamma'} u_{R \alpha} s_{R \beta} s_{L \gamma} u_{R \alpha'} 
s_{R \beta'} s_{L \gamma'},
\end{equation}
where the color indices and chiralities of the quarks are noted, 
the contribution of Eq.~(12) is then given by\cite{15}
\begin{equation}
T_1 \simeq {{g^4 \lambda^2_{usb} A^2 m_b^2 m_{\tilde w}} \over {32 \pi^2 
\tilde m^8}} V_{ts}^2 F(m^2_{\tilde w}, M_W^2, \tilde m^2, m_t, A),
\end{equation}
where we have assumed universal soft supersymmetry-breaking scalar masses, 
$V_{ts}$ is the quark-mixing entry for $t$ to $s$ through the $W$ boson, and
\begin{eqnarray}
F &=& {1 \over 2} J(m^2_{\tilde w}, M_W^2, \tilde m^2 + m_t^2 + A m_t, m_t^2) 
+ {1 \over 2} J(m^2_{\tilde w}, M_W^2, \tilde m^2 + m_t^2 - A m_t, m_t^2) 
\nonumber \\ &-& J(m^2_{\tilde w}, M_W^2, \tilde m^2, m_t^2) - \{ m_t^2 
\rightarrow 0 ~{\rm in~the~last~entry~of~each~term} \},
\end{eqnarray}
with the function $J$ given by\cite{10,13}
\begin{equation}
J(a_1, a_2, a_3, a_4) = \sum_{i=1}^4 {{a_i^2 \ln (a_i)} \over {\prod_{k \neq i} 
(a_i - a_k)}}.
\end{equation}
Using the correspondence of $n - \bar n$ oscillation to $N N$ annihilation 
inside a nucleus, we estimate the lifetime of strangelets from the above 
effective interaction assuming $A = 200$ GeV, $\lambda_{usb} < 1$, and 
$\tilde m > 200$ GeV, to be
\begin{equation}
\tau > 10^{22}{\rm y}.
\end{equation}
This tells us that such contributions are negligible from the supersymmetric 
standard model.

Recently, another contribution to Eq.~(2) has been identified\cite{13} 
involving $\lambda_{tds}$ and $\lambda_{tdb}$ without the GIM suppression 
of Eq.~(16).  The form of its contribution to Eq.~(7) is
\begin{equation}
{\cal H}_{int} = T_2 \epsilon_{\alpha \beta \gamma} \epsilon_{\alpha' 
\beta' \gamma'} u_{L \alpha} s_{L \beta} s_{R \gamma} u_{L \alpha'} 
s_{L \beta'} s_{R \gamma'},
\end{equation}
which is not the same as Eq.~(15).  The important coupling is now 
$\lambda_{tsb}$, but $T_2$ is also suppressed relative to $T_1$ by 
$V_{ub}^2$, hence this contribution to Eq.~(7) is even more negligible.

\section{Extended Models Involving Strangeness}

Although the supersymmetric standard model cannot have a sizable contribution 
to the $\Delta B = 2$, $\Delta N_s = 4$ operator of Eq.~(7), an extended model 
including additional particles belonging to the fundamental {\bf 27} 
representation of $E_6$, inspired by superstring theory\cite{16}, may do 
better.  Consider a slight variation of the model of exotic baryon-number 
nonconservation proposed some years ago\cite{17}.  In addition to the 
usual quark and lepton superfields
\begin{equation}
Q \sim (3, 2, 1/6), ~~~ u^c \sim (\bar 3, 1, -2/3), ~~~ d^c \sim (\bar 3, 
1, 1/3);
\end{equation}
\begin{equation}
L \sim (1, 2, -1/2), ~~~ e^c \sim (1, 1, 1), ~~~ N^c \sim (1, 1, 0);
\end{equation}
we have
\begin{equation}
h \sim (3, 1, -1/3), ~~~ h^c \sim (\bar 3, 1, 1/3);
\end{equation}
\begin{equation}
E \sim (1, 2, -1/2), ~~~ \bar E \sim (1, 2, 1/2), ~~~ S \sim (1, 1, 0);
\end{equation}
each transforming under the standard $SU(3) \times SU(2) \times U(1)$ gauge 
group as indicated.  Imposing the discrete symmetry $Z_2 \times Z_2$ under 
which
\begin{eqnarray}
Q, u^c, d^c, N^c &\sim& (+, -), \\ L, e^c &\sim& (-, +), \\ h, h^c, E, \bar E, 
S &\sim& (+, +),
\end{eqnarray}
we get the allowed terms
\begin{equation}
Q Q h, ~~ u^c d^c h^c, ~~ d^c h N^c, ~~ N^c N^c,
\end{equation}
in the superpotential, but not $d^c h$.  We also assume that $\tilde N^c$ 
does not develop a vacuum expectation value, so that $B$ is broken 
explicitly by the above $N^c N^c$ term only.  [Note that without this last 
term, we can assign $B = -2/3, 2/3, 1$ to $h, h^c, N^c$ respectively and 
$B$ would be conserved.]  This model allows us to obtain an effective 
$(uss)^2$ operator without going through a loop, as shown in Fig.~1.

Using the formalism of Ref.~[10] for the process
\begin{equation}
(A, N_s) \rightarrow (A-2, N_s-2) + KK,
\end{equation}
we estimate the lifetime to be
\begin{equation}
\tau \sim {{32 \pi m_N^2} \over {9 \rho_N}} \left[ {M \over \tilde \Lambda} 
\right]^{10} \sim 1.2 \times 10^{-28} \left[ {M \over \tilde \Lambda} 
\right]^{10} {\rm y},
\end{equation}
where $m_N \sim 1$ GeV, $\rho_N \sim 0.25$ fm$^{-3}$ is the nuclear density, 
and $\tilde \Lambda \sim 0.3$ GeV is the effective interaction energy scale 
corresponding to Eq.~(5).  If we assume $M = 1$ TeV, then
\begin{equation}
\tau \sim 2 \times 10^7 {\rm y}.
\end{equation}
In the above, we have assumed that $\lambda_{ush}$ is unconstrained 
phenomenologically.  However, consider the term  $Q_1 Q_2 h$ 
where $Q_1 = (u, d')$ and $Q_2 = (c, s')$ so that $Q_1 Q_2 = u s' - c d'$.  
Hence from this term alone, $\lambda_{udh} = (V_{cd}/V_{cs}) \lambda_{ush}$. 
Since $\lambda_{udh}$ may now combine with $\lambda_{s^c h N^c}$ to induce 
$N N \rightarrow K K$ decays, its magnitude is seriously suppressed and 
$\lambda_{ush}$ would be too small for Eq.~(31) to be valid.  However, there 
is also a third generation, allowing us the $Q_1 Q_3 = u b' - t d'$ term 
which may then be fine-tuned to eliminate the $\lambda_{udh}$ component, thus 
saving Eq.~(31).  This is of course not a very natural solution, but cannot 
otherwise be ruled out.

We may also consider the following tailor-made extension.  Let there be a 
new exotic scalar multiplet $\tilde Q_6 \sim (\bar 6, 1, 2/3)$ with 
$B = -2/3$ and a new exotic fermion multiplet $\psi_8 \sim (8, 1, 0)$ 
with $B = 1$.  Assume the existence of Yukawa terms $s^c s^c \tilde Q_6^*$, 
$u^c \psi_8 \tilde Q_6$, and the $B$-nonconserving Majorana mass term 
$\psi_8 \psi_8$, then an effective $(uss)^2$ term is possible, as shown in 
Fig.~2.  Note that the $uss$ combination is now a color octet.  This may 
in fact be a more efficient way to dissipate strange matter which is 
presumably not clumped into color-singlet constituents as in ordinary nuclei.

\section{Stability Limit of Strange Matter}

Since $\tau$ depends on $M/\tilde \Lambda$ to the power 10 in Eq.~(30), it 
appears that a much shorter lifetime than that of Eq.~(31) is theoretically 
possible.  However, there is a crucial phenomenological constraint which 
we have yet to consider.  Although two nucleons cannot annihilate inside 
a nucleus to produce four kaons, they can make three kaons plus a pion. 
The effective $(uss)^2$ operator must now be supplemented by a weak 
transition $s \rightarrow u + d + \bar u$.  We 
can compare the effect of this on $N N \rightarrow K K K \pi$ versus that 
of the $(uds)^2$ operator on $N N \rightarrow K K$ discussed in Ref.~[10]. 
First let us look at the phase-space difference.  Consider a nucleus with 
atomic number $A$ decaying into one with atomic number $A-2$ plus two kaons 
with energy-momentum conservation given by $p = p' + k_1 + k_2$.  The decay 
rate is proportional to
\begin{eqnarray}
F_1 &=& {1 \over {(2 \pi)^5}} \int {{d^3 k_1} \over {2 E_1}} \int {{d^3 k_2} 
\over {2 E_2}} \int {{d^3 p'} \over {2 E'}} \delta^4 (p - p' - k_1 - k_2) 
\nonumber \\ &\simeq& {1 \over {2 M'}} {1 \over {(2 \pi)^5}} \int {{d^3 k_1} 
\over {2 E_1}} \int {{d^3 k_2} \over {2 E_2}} \delta (M - M' - E_1 - E_2) 
\nonumber \\ &\simeq& {1 \over {2 M'}} {1 \over {(2 \pi)^3}} 
\int_{m_K}^{E_{max}} k_1 k_2 dE_1,
\end{eqnarray}
where $k_1 = (E_1^2 - m_K^2)^{1/2}$, $k_2 = [(2 m_N - E_1)^2 - m_K^2]^{1/2}$, 
and $E_{max} = 2 m_N - m_K$.  
\newpage
Consider next $A \rightarrow (A-2) + K K K \pi$ 
with $p = p' + k_1 + k_2 + k_3 + k_\pi$.  Because of the limited phase 
space, we assume the kaons to be nonrelativistic.  Hence
\begin{eqnarray}
F_2 &\simeq& {1 \over {2 M'}} {1 \over {(2 \pi)^{11}}} \int {{d^3 k_1} \over 
{2 m_K}} \int {{d^3 k_2} \over {2 m_K}} \int {{d^3 k_3} \over {2 m_K}} \int 
{{d^3 k_\pi} \over {2 E_\pi}} \delta (M - M'- 3 m_K - {{k_1^2 + k_2^2 + 
k_3^2} \over {2 m_K}} - E_\pi) \nonumber \\ &\simeq& {1 \over {2 M'}} 
{1 \over {(2 \pi)^7}} {1 \over m_K^3} \int k_\pi k_1^2 k_2^2 k_3^2 dk_1 dk_2 
dk_3,
\end{eqnarray}
where $k_\pi = \{[2m_N - 3m_K - (k_1^2 + k_2^2 + k_3^2)/2m_k]^2 - m_\pi^2
\}^{1/2}$.  The above integral can be evaluated by treating $k_{1,2,3}$ as 
Cartesian coordinates and then convert them to three-dimensional polar 
coordinates.  The angular integration over the $k_{1,2,3} > 0$ octant 
yields a factor of $\pi/210$ and we get
\begin{equation}
F_2 \simeq {1 \over {2 M'}} {1 \over {(2 \pi)^7}} {1 \over m_K^3} {\pi \over 
{210}} \int_0^{k_{max}} k_\pi k^8 dk,
\end{equation}
where $k_{max} = [2 m_K (2 m_N - 3 m_K - m_\pi)]^{1/2}$.  The effective 
interaction here also differs from that of Eq.~(32) by the appearance of 
a third kaon and an extra pion which can be thought of as having been 
converted by the weak interaction from a fourth kaon.  We thus estimate the 
suppression factor to be
\begin{equation}
\xi \sim \left[ {{f_K^2 T_{K \pi}} \over {\tilde \Lambda^6}} \right]^2 
{F_2 \over F_1} \sim 10^{-20},
\end{equation}
where $f_K = 160$ MeV, and $T_{K \pi} = 0.07$ MeV$^2$ is obtained using the 
symmetric soft-pion reduction\cite{18} of the experimental 
$K \rightarrow 2 \pi$ amplitude.  We have again invoked the effective 
interaction energy scale $\tilde \Lambda = 0.3$ GeV used in Eq.~(30).  
Obviously, our estimate depends very sensitively on this parameter, but 
that is not untypical of many calculations in nuclear physics.  Since the 
stability of nuclei is at least $1.6 \times 10^{25}$ years, the reduction 
by $\xi$ of the above yields
\begin{equation}
\tau > 10^5 {\rm y}
\end{equation}
for the lifetime of strangelets against $\Delta B = 2$, $\Delta N_s = 4$ 
decays.

\section{Concluding Remarks}

We have pointed out in this paper that an effective $(uss)^2$ interaction 
may cause stable strange matter to decay, but the lifetime is constrained 
phenomenologically by the stability of nuclei against decays of the type 
$A \rightarrow (A-2) + K K K \pi$ and we estimate it to have a lower limit 
of $10^5$ years.  This result has no bearing on whether stable or metastable 
strangelets can be created and observed in the laboratory, but may be 
important for understanding whether there is stable strange bulk matter 
left in the Universe after the Big Bang and how it should be searched for. 
For example, instead of the usual radioactivity of unstable nuclei, strange 
matter may be long-lived kaon and pion emitters.  The propagation of these 
kaons and pions through the matter itself and their interactions 
within such an environment are further areas of possible study.  Since 
energy is released in each such decay, although the amount is rather 
small, it may be sufficient to cause a chain reaction and break up the 
bulk matter in a cosmological time scale.  Furthermore, if the particles 
mediating this effective interaction have masses of order 1 TeV as 
discussed, then forthcoming future high-energy accelerators such as the 
Large Hadron Collider (LHC) at the European Center for Nuclear Research 
(CERN) will have a chance of confirming or refuting their existence.
\vspace{0.3in}
\begin{center} {ACKNOWLEDGEMENT}
\end{center}

This work was supported in part by the U. S. Department of Energy under 
Grant No. DE-FG03-94ER40837.

\newpage
\bibliographystyle{unsrt}

\begin{thebibliography}{99}
\bibitem{1} E. Witten, Phys. Rev. {\bf D30}, 272 (1984).
\bibitem{2} S. Chin and A. Kerman, Phys. Rev. Lett. {\bf 43}, 1292 (1979).
\bibitem{3} Particle Data Group, R. M. Barnett {\it et al.}, Phys. Rev. 
{\bf D54}, 1 (1996).
\bibitem{4} Particle Data Group, L. Montanet {\it et al.}, Phys. Rev. 
{\bf D50}, 1173 (1994).
\bibitem{5} M. Baldo-Ceolin {\it et al.}, Z. Phys. {\bf C63}, 409 (1994).
\bibitem{6} Ch. Berger {\it et al.}, Phys. Lett. {\bf B269}, 227 (1991).
\bibitem{7} C. B. Dover, A. Gal, and J. M. Richard, Phys. Rev. {\bf C31}, 
1423 (1985); Phys. Lett. {\bf B344}, 433 (1995).
\bibitem{8} See for example T. K. Kuo and S. T. Love, Phys. Rev. Lett. 
{\bf 45}, 93 (1980).
\bibitem{9} R. Barbieri and A. Masiero, Nucl. Phys. {\bf B267}, 679 (1986); 
S. Dimopoulos and L. J. Hall, Phys. Lett. {\bf B196}, 135 (1987).
\bibitem{10} J. L. Goity and M. Sher, Phys. Lett. {\bf B346}, 69 (1995).
\bibitem{11} See for example J. Madsen, in ``Strangeness in Hadronic Matter",
ed. by J. Rafelski (American Institute of Physics, New York, 1995), p.~32.
\bibitem{12} A partial list includes B. Brahmachari and P. Roy, Phys. Rev. 
{\bf D50}, R39 (1994); 
G. Bhattacharyya, D. Choudhury, and K. Sridhar, Phys. Lett. {\bf B355}, 193 
(1995); C. E. Carlson, P. Roy, and M. Sher, Phys. Lett. {\bf357}, 99 (1995); 
A. Yu. Smirnov and F. Vissani, Nucl. Phys. {\bf B460}, 37 (1996); 
Phys. Lett. {\bf B380}, 317 (1996).
\bibitem{13} D. Chang and W.-Y. Keung, hep-ph/9608313 (August 1996).
\bibitem{14} S. L. Glashow, J. Iliopoulos, and L. Maiani, Phys. Rev. 
{\bf D2}, 1285 (1970).
\bibitem{15} Equation (3) of Ref.~[10] is missing a factor of 12 in the 
denominator: J. Goity and M. Sher, private communication.
\bibitem{16} P. Candelas, G. T. Horowitz, A. Strominger, and E. Witten, 
Nucl. Phys. {\bf B258}, 46 (1985).
\bibitem{17} E. Ma, Phys. Rev. Lett. {\bf 60}, 1363 (1988).
\bibitem{18} M. Suzuki, Phys. Rev. {\bf 144}, 1154 (1966).
\end{thebibliography}

~~~
\begin{center} {\large \bf Figure Captions}
\end{center}

\noindent Fig.~1.  Effective $(uss)^2$ interaction from a supersymmetric 
model with $E_6$ particle content.

\noindent Fig.~2.  Effective $(uss)^2$ interaction from a model with 
exotic color-sextet scalar and color-octet fermion.

\end{document}